\documentclass[%
 reprint,
showpacs,
amsmath, amssymb,
aps,
prl,
]{revtex4-1}

\usepackage{graphicx}
\usepackage{dcolumn}
\usepackage{bm}
\usepackage{color}

\newcommand{\units}[1]{\:\mathrm{#1}}            
\newcommand{\idx}[1]{_{\mathrm{#1}}}          

\begin{document}


\title{Spatial separation in a thermal mixture of ultracold $^{174}$Yb and $^{87}$Rb atoms}

\author{F.~Baumer}
\author{F.~M\"{u}nchow}
\author{A.~G\"{o}rlitz}
\email{axel.goerlitz@uni-duesseldorf.de}
\affiliation{Institut f\"{u}r Experimentalphysik,
Heinrich-Heine-Universit\"{a}t D\"{u}sseldorf, Universit\"{a}tsstra{\ss}e 1,
40225~D\"{u}sseldorf, Germany}
\author{S.~E.~Maxwell}
\author{P.~S.~Julienne}
\author{E.~Tiesinga}

\affiliation{Joint Quantum Institute and the National Institute of Standards and Technology, Gaithersburg, MD, USA, 20899}


\begin{abstract}
We report on the observation of unusually strong interactions in a thermal mixture of ultracold atoms which cause a significant modification of the spatial distribution. A mixture of $^{87}$Rb and $^{174}$Yb with a temperature of a few $\mu$K is prepared in  a hybrid trap consisting of a bichromatic optical potential superimposed on a magnetic trap. For suitable trap parameters and temperatures, a spatial separation of the two species is observed. 
We infer that the separation is driven by a large interaction strength between $^{174}$Yb and $^{87}$Rb accompanied by a large three-body recombination rate. Based on this assumption we have developed a diffusion model which reproduces our observations.

\end{abstract}

                                
\maketitle

In collisions of ultracold atoms, the strength of the interparticle interaction
is determined by subtle details of the interatomic potential \cite{Chin2010}. It is impossible in practice to predict the behavior of an ultracold sample {\it ab initio} and the experimental investigation of a single- or multi-species ensemble of ultracold atoms may be full of surprises. A prominent example is $^{133}$Cs where after years of experimental studies it became clear that without manipulation using external magnetic fields \cite{vuletic1999,weber2003} the scattering length is extremely large due to a zero-energy resonance \cite{arndt1997}.

A fundamental question related to the interaction between atomic species is whether different components may occupy the same regions in space or separate spatially. In trapped quantum degenerate gases this question of miscibility is determined by the interplay between single-species and interspecies interactions \cite{pu1998a}. For such systems, phase separation has  been observed in a dual species Bose-Einstein condensate with $^{87}$Rb and $^{85}$Rb \cite{Papp2008} as well as in a strongly interacting Bose-Fermi mixture of Li$_2$-molecules and $^6$Li atoms \cite{Shin2008}. 

In a dual-component gas of thermal atoms, the situation is somewhat different because the thermal motion, which counteracts  interaction-driven spatial separation, has to be taken into account. However, for collisionally dense samples with a  mean-free path much smaller than the sample size, motion is diffusive and together with a loss process with a spatial and/or density dependence this may also lead to a spatial separation of components. Such a process, for example, contributes to the stratification of chemical elements inside stars \cite{Alecian2006}. 

In this Letter, we report on the observation of spatial separation in an ultracold thermal mixture of co-trapped $^{87}$Rb and $^{174}$Yb atoms, indicating large interspecies interactions.
The spatial separation is detected as an exclusion of the Yb atoms from the region of high Rb density at temperatures  of a few $\mu$K.

\begin{figure}
\includegraphics{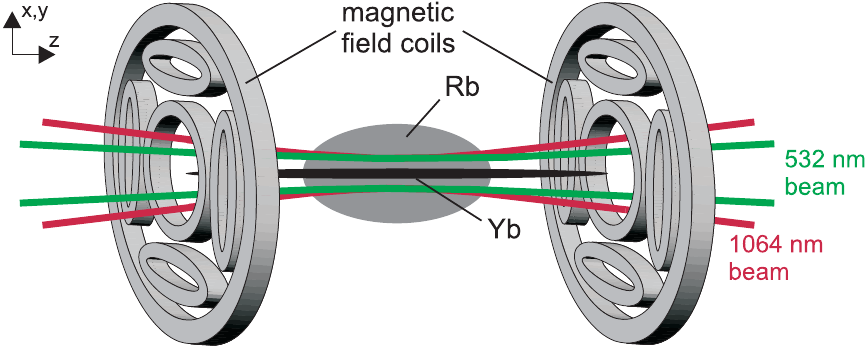}
\caption{\label{fig:combtrap}(Color online) Sketch of the trapping geometry for $^{174}$Yb and $^{87}$Rb showing the trap lasers and magnetic field coils (not to scale). Coordinate axes $x,y,z$ are indicated.}
\end{figure}

The trapping potential for the two atomic species is composed of a clover-leaf magnetic trap (MT) and a bichromatic optical dipole trap (BIODT), as shown in Fig.~\ref{fig:combtrap}.  The diamagnetic Yb atoms are only confined by the BIODT created by two superimposed  light fields  with wavelengths of $532\units{nm}$ and $1064\units{nm}$.   Both  fields are red-detuned with respect to the dominant  transition in Yb at $399\units{nm}$ \cite{Tassy2010}. 
The total potential for Rb, in contrast, is created by the MT, and the repulsive (blue-detuned with respect to the dominant $780\units{nm}$ transition) $532\units{nm}$ light field plus the attractive (red-detuned) $1064\units{nm}$ field \cite{Onofrio2002}.  Thus, the optical potential for Yb is always attractive, while it can be tuned from repulsive to attractive for Rb by changing the powers in the two light fields. The MT and the BIODT have their weak axes aligned along the $z$-direction for maximum overlap of the two atomic clouds. A related scheme was recently used  in Ref. \cite{Catani2009}.

In a typical experimental sequence, Yb atoms are loaded from a magneto-optical trap, operating on the $^1$S$_0$ $\rightarrow$ $^3$P$_1$ transition at $556\units{nm}$ \cite{nemitz2009}, into the BIODT. The BIODT light fields 
have beam waists of $w_{1064} \approx 15\units{\mu m}$ and $ w_{532} \approx 16\units{\mu m}$ with initial power levels of $P_{1064} = 110\units{mW}$ and $P_{532} = 3.4\units{W}$. The relative position of the two beams is actively stabilized.  
 Next, the power of the $532\units{nm}$  beam is ramped over $30\units{s}$ to a final level of $P_{532} = 340\units{mW}$. After this ramp, the Yb atom number has decreased to $\approx2\times10^5$ and the Yb temperature is $\approx5\units{\mu K}$. In this configuration, the trap frequencies are $(1.05\pm0.02)\units{kHz}$ radially and $ (9.5\pm0.2)\units{Hz}$ axially and the calculated trap depth is $U/k\idx{B} \approx 60\units{\mu K}$, where  $k_{\rm B}$ is the Boltzmann constant. If not stated otherwise, all errors represent one standard deviation combined statistical and systematic uncertainty throughout this manuscript. 

While the potential for Yb is ramped to its final value, an ensemble of $10^7$  Rb atoms is prepared in a MT in the ($F=1, m_F=-1$) state at a temperature of $\approx1.5\units{\mu K}$. 
The  Rb peak density in the MT is a few $10^{13}\units{cm}^{-3}$. The center of the MT is initially located $0.7\units{mm}$ from the BIODT and the Yb cloud is almost unaffected by the Rb preparation process, which takes $35\units{s}$.

Subsequently, the MT is moved to the position of the BIODT within $500\units{ms}$ by applying a magnetic bias. At this stage, the effect of the BIODT on the Rb is minimal. As the two atomic species make contact, we observe rapid thermalization of Yb with the colder Rb cloud to the Rb temperature.  Heating of the Rb cloud is negligible due to the large difference in atom numbers.

The Rb cloud is then compressed by increasing the power of the $1064$\,nm beam within $100\units{ms}$, thus making the BIODT attractive for Rb. For $P_{1064} = 140\units{mW}$ the measured trap frequencies for  Rb are $(800\pm40)\units{Hz}$ radially and $ (24\pm1)\units{Hz}$ axially. The Rb temperature increases by adiabatic compression to $2\,-\,3\units{\mu K}$ and the Rb peak density rises up to a maximum value of $3\times 10^{14}\units{cm}^{-3}$. 
We have confirmed experimentally that the two gases are in thermal equilibrium  after  deformation of the potential, and define a temperature $T_{\rm Yb}=T_{\rm Rb} \equiv T$.
The deformation changes the total potential depth for Yb by less than 10\%. No signs of quantum degeneracy are observed.

The experimentally determined Rb cloud shapes  are consistent
with a model calculation based on a Gaussian beam profile, measured temperatures and trap
frequencies, magnetic moments, and atomic polarizabilities. Densities are
assumed to follow a Boltzmann distribution in the trap potential.
Though the qualitative shape is reproduced, the modeling of density is limited by the strong dependence of the density on uncertainties in the potential.  The distribution of the Yb when the Rb cloud is not present is consistent with a similar model calculation.

\begin{figure}
\includegraphics[width=1\columnwidth]{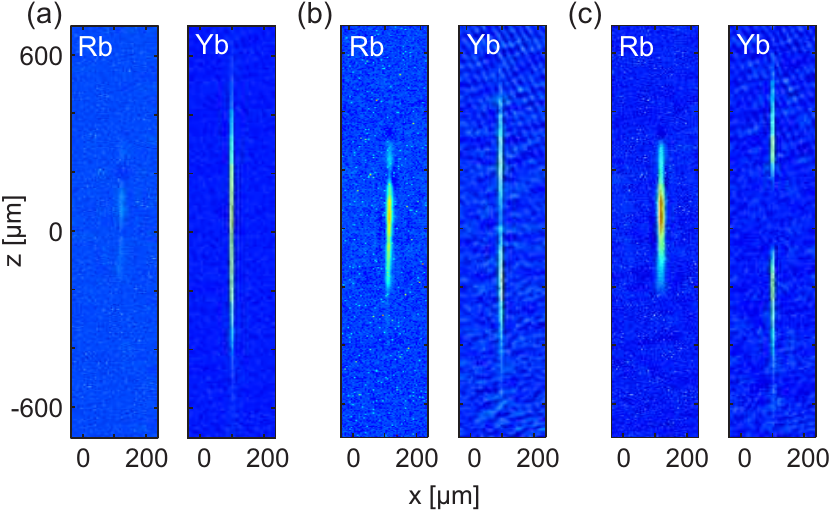}
\caption{\label{fig:rawdata}(Color online) False color {\it in situ} images of $^{87}$Rb and $^{174}$Yb clouds at  $T\approx3\units{\mu K}$. For $^{174}$Yb we use absorption imaging while for $^{87}$Rb we use dark-contrast imaging \cite{Ketterle1999b}.  The $^{87}$Rb peak density increases from  (a) $1.4\times10^{13}\units{cm^{-3}}$   to (b) $7.3\times10^{13}\units{cm^{-3}}$  and to (c) $1.4\times10^{14}\units{cm^{-3}}$. In the radial dimension the images are resolution limited.}
\end{figure}

When the compression, and hence the  density of the Rb cloud is large, we observe a spatial separation of the two atomic clouds, with the Yb bracketing the Rb cloud. In Fig.~\ref{fig:rawdata}, simultaneously taken images of both species are shown.   Additional time-of-flight images of Rb are used to determine absolute atom number and temperature. The extent to which the two species separate depends on the number of Rb atoms, the density distribution and the temperature.  We do not observe a distortion of the Rb cloud caused by the presence of the Yb. Any such effect is expected to be small because of the  small  Yb density $n\idx{Yb}\approx10^{13}\units{cm}^{-3}$. The observed demixing is accompanied by a rapid loss of Yb atoms, which is attributed to inelastic collisions.  No significant loss of Rb atoms is observed. A 1/e-lifetime of $\approx 200\units{ms}$ of the Yb cloud has been measured  for an initial  Rb peak density $n_{\mathrm{Rb}}(\vec 0) = (3\pm1.8)\times 10^{14}\units{cm}^{-3}$. We attribute the loss  to three-body recombination of Yb+Rb+Rb, which is expected to become important when the interspecies scattering is strong \cite{Fedichev1996b,Kraemer2006,Esry1999}. Two-body loss processes do not exist in this system due to the lack of structure in the spherically symmetric Yb ground state.  Although we have observed a spatial separation with the MT switched both on and off, all results described in this Letter have been obtained with the MT switched on.

We take two approaches to quantify the spatial separation.  First, we attempt to explain the phenomenon in terms of a mean field interaction.  Second, we model the system via a diffusive model that includes three-body recombination.

The mean-field approach leads us to quantify the spatial separation  of Yb and Rb in terms of an additional potential for Yb due to the elastic interaction with Rb atoms. Formally, the total potential seen by the Yb atoms is taken to be
\begin{equation}
\label{eq:UtotYb}
U_{\mathrm{Yb}}(\vec{r})=V(\vec{r})+{U}_{\rm YbRb}\, n_{\mathrm{Rb}}(\vec{r}) ,
\end{equation}
where $V(\vec{r})$ is the trap potential due to the BIODT and the interaction parameter ${U}_{\rm YbRb}$ quantifies the interspecies interaction. Both  $U_{\rm YbRb}$ and $n_{\rm Rb}$ can depend on temperature.  Effects of the interspecies interaction  on the Rb density distribution are neglected because of the comparatively low density of Yb.  

\begin{figure}
\includegraphics[width=1\columnwidth]{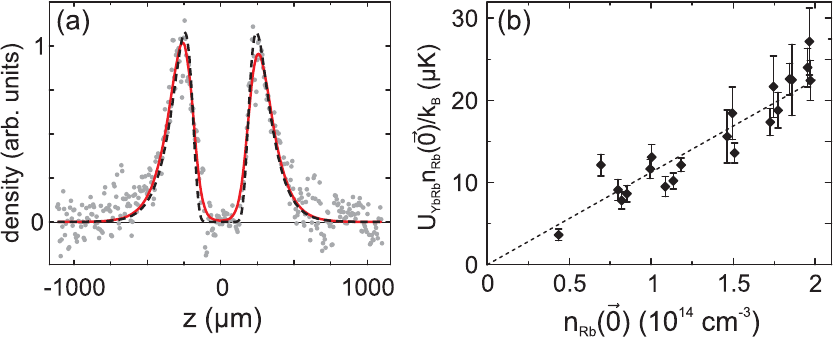}
\caption{\label{fig:atomnum}(Color online) (a) Typical axial $^{174}$Yb density  distribution for $T = (2.5 \pm 0.3) \,\mu$K  and $n_{\mathrm{Rb}}(\vec 0) = (3\pm 1.8)\times10^{14}\units{cm^{-3}}$. The solid red curve is a fit of the $^{174}$Yb density distribution using a mean field model while the dashed black curve is from a diffusion model. (b) Fitted value of ${U}_{\rm YbRb}n_{\rm Rb}(\vec 0)$ using a mean field model as a function of $n_{\mathrm{Rb}}(\vec 0)$ for measurements  at $T = 3.2\units{\mu K}$.  The error bars represent only fit uncertainties.}
\end{figure}

We fit Eq.~\ref{eq:UtotYb} to the experimental data assuming that the density of Yb follows a Boltzmann distribution, $e^{-U_{\mathrm{Yb}}(\vec{r})/k_{\rm B}T}$.   Besides the relative trap center position, ${U}_{\rm YbRb}$ is the sole fit parameter because the Rb density distribution and the temperature are determined experimentally.  A sample fit is shown in Fig.~\ref{fig:atomnum}\,(a).
  Using this method, we obtain ${U}_{\rm YbRb}/k_{\rm B}=(9.7\pm 5.2) \times 10^{-14}\units{\mu K\, cm^3}$  from a series of measurements at $T = 3.2\units{\mu K}$, as shown in Fig.~\ref{fig:atomnum}\,(b).  A spatial separation of Yb and Rb was observed between $1.5\units{\mu K}$ and $6.5\units{\mu K}$. Within experimental uncertainties, this analysis results in  a temperature-independent value for $U_{\mathrm{YbRb}}$.

In a standard mean-field approach ${U}_{\rm YbRb}$ is related to a characteristic length $a_{\mathrm{YbRb}}$ by ${U}_{\rm YbRb}\,=\,2 \pi \hbar^2 a_{\mathrm{YbRb}}/{\mu}$,  
\cite{Blatt1952, Huang1957} where $\mu$ is the reduced mass of $^{87}$Rb + $^{174}$Yb.  Using this relationship we obtain a value of $a_{\mathrm{YbRb}} = 35000 \pm 19000 a_0$, with $a_0$ the Bohr radius.   One is tempted to take this value as a measurement of the zero-energy interspecies scattering length since the system is well below the $p$-wave barrier of $60\units{\mu K}$~\cite{Tassy2010, Landau1989}.  However, for large scattering lengths the system is well outside of the energy independent regime, and such an interpretation is invalid.
 A next interpretation would be that the length is the thermal average of an energy dependent $s$-wave scattering length~\cite{Huang1957}.    We use the effective-range theory, which provides an accurate representation of energy dependence at large scattering lengths \cite{Blume2002,Gao1998}. Thermal averaging for any zero-energy scattering length yields a value for $a_{\mathrm{YbRb}}$ that is much smaller than the measured length. Thus, we conclude that taking the length  to be a scattering length or a thermally averaged scattering length is incompatible with theory. Nevertheless, the fitted values of ${U}_{\rm YbRb}n_{\rm Rb}(\vec 0)$ shown in Fig. \ref{fig:atomnum}(b) are a measure for the total strength of the interaction and clearly demonstrate its dependence on the Rb density $n_{\mathrm{Rb}}(\vec{r})$.

While mean-field theory does not provide a satisfactory explanation, the results do suggest that the interactions between Rb and Yb are  strong and that the elastic cross section is large.  Correspondingly, the mean free path of a thermal Yb atom in a cloud of Rb will be much smaller than the size of the cloud.  Thus we can model the trapped system using the time-dependent diffusion equation including loss due to three-body recombination.  
The diffusion equation is  
\begin{equation}
 \dot n({\bf r},t)=-\nabla\cdot{\bf j}({\bf r},t)-K_{3}n^{2}_{\rm  Rb}({\bf r})  n({\bf r},t),
\end{equation}
where  $n({\bf r},t)$ is the Yb distribution, $K_{3}$ is the three-body loss rate coefficient, $n_{\rm  Rb}({\bf r})$ is the Rb density, which will be assumed to follow a Boltzmann distribution and is independent of time,  and the Yb current density is
\begin{equation}
{\bf j}({\bf r},t)=-\frac{k_{B}T\tau({\bf r})}{m}\nabla n({\bf r},t)+
\frac{\tau({\bf r})}{m}\left(\nabla V({\bf r})\right) n({\bf r},t),
\end{equation}
which describes diffusion and drift.
Here $\tau({\bf r})$ is the spatially dependent mean collision time and $m$ is the $^{174}$Yb mass. 

The mean collision time is determined by Yb+Yb and Rb+Yb collisions and satisfies $\tau({\bf r})^{-1}=\tau_{\rm Yb Rb}({\bf r})^{-1}+\tau_{\rm YbYb}({\bf r})^{-1}$.
 Here $\tau_{\rm Yb  Rb}({\bf r})^{-1}=
\left< v_{\rm rel} \sigma_{\rm YbRb} \right>n_{\rm  Rb}({\bf r})$, where $\sigma_{\rm YbRb}$ is the interspecies collision cross section, and $v_{\rm rel}$ is the relative velocity.  The angle brackets represent a thermal average.  Similarly, self-diffusion is governed by $\tau_{\rm Yb Yb}({\bf r})^{-1}=
\left< v_{\rm rel,Yb} \sigma_{\rm YbYb} \right>n({\bf r})$. 

By looking for the solution $n({\bf r},t)=e^{-\Gamma t}n({\bf r})$, the diffusion equation becomes an eigenvalue equation for the loss coefficient $\Gamma$.  The smallest eigenvalue corresponds to the longest lived eigenfunction. This eigenfunction is the spatial distribution that would be observed in an experiment at long times.  For $K_{3}=0$ in a trap that is deep compared to the temperature, this solution is a Boltzmann distribution, $n_{\rm Yb}\propto e^{-V({\bf r})/k_{B}T}$, with $\Gamma=0$.

  Rather than solving the full 3-D  equation including the self-diffusion, which depends on the Yb density and is nonlinear, we choose to reduce the equation to 1-D by making a few simplifying assumptions and then integrating out the transverse coordinates.  The reduction is justified by the cylindrical symmetry of the cigar-shaped trapping potential, taken to be harmonic with no axial dependence in the transverse direction, and the larger radial extent of the Rb cloud.   
As one may assume $\tau_{\rm Yb Yb}({\bf r}) \gg \tau_{\rm YbRb}({\bf r})$, we substitute the real spatial dependence of $\tau_{\rm Yb Yb}({\bf r})$ by a functional form that turns $\tau({\bf r})$ into a separable function of radial and axial coordinates and removes the nonlinearity. With the known Yb+Yb cross section $\sigma_{\rm YbYb}=7.7\times 10^{-12}$~cm$^2$ \cite{Kitagawa2008}, which is constant over the range of energies here, $\tau_{\rm Yb Yb}= 5.2$~ms  for an Yb density of $10^{13}$~cm$^{-3}$.  
 After the trap geometry has been set, the adjustable inputs to the theory are $\left<v_{\rm rel}\sigma_{\rm YbRb}\right>$,  $K_3$, and $\tau_{\rm Yb Yb}({\bf r})$. Because the interactions appear to be stronger than those seen in any other ultracold thermal gas, we initially perform a model calculation of diffusion using the unitarity limit,  $\sigma_{\rm YbRb}=4\pi\hbar^2/(\mu v_{rel})^2$.  In this limit, $\tau_{\rm YbRb}$ can be computed analytically. For values of  $T=2.5\,\mu$K and $n_{Rb}(0)=3\times10^{14}$cm$^{-3}$, corresponding to the experimental parameters of Fig.~\ref{fig:atomnum}\,(a), one obtains $\tau_{\rm Yb Rb}(\vec 0)= 5.3\,\mu$s. The unitarity-limited value for $K_{3}$ is proportional to $T^{-2}$ and at $2.5\units{\mu K}$ it has a value of $8\times 10^{-25}{~\rm cm}^{6}{\rm s}^{-1}$ \cite{Esry2008}.  

\begin{figure}
\includegraphics[width=1\columnwidth]{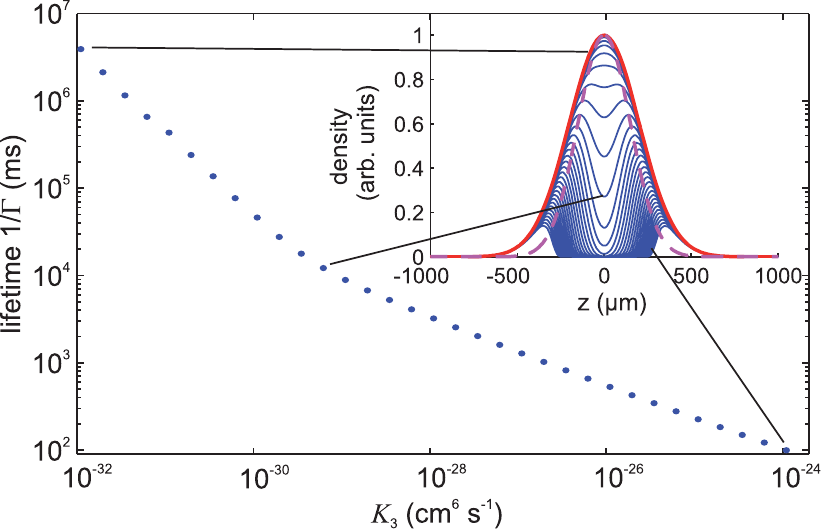}\caption{\label{fig:fullfigure}(Color online) Model calculation of the asymptotic lifetime of trapped $^{174}$Yb versus the Rb+Rb+Yb three-body loss rate coefficient, $K_3$, assuming a unitarity-limited scattering cross section.  Inset shows the axial profile of  $^{174}$Yb  for varying $K_3$ (thin blue).  Boltzmann distributions for $^{174}$Yb (thick red) and $^{87}$Rb (dashed pink) are given.  Lines connect key points on the lifetime plot to the corresponding distribution in the inset. }
\end{figure}

Figure~\ref{fig:fullfigure} shows 1-D solutions in a harmonic trap, giving the lifetime $1/\Gamma$  and profile of the longest lived eigenfunction for varying $K_3$.  For small $K_3$, the Yb distribution is Gaussian and $\Gamma\propto K_3$.  With increasing $K_3$, a kink in the lifetime occurs where the Yb is excluded from the Rb.  The reduced overlap of the clouds results in a weaker dependence  of lifetime on $K_3$.  The next longest-lived symmetric diffusion mode has a lifetime that is nearly ten times shorter.

Though it contains significant simplifications, the 1-D diffusion model
reproduces the experimental density profile as shown in the
dashed black curve Fig.~\ref{fig:atomnum}\,(a). The curve is calculated using
the model of the axial potential that is used in the mean field calculations
and the observed $T$ and $n_{\rm Rb}(\vec{0})$. The potential has weaker
confinement away from the center than the harmonic approximation. 
Both $\left<v_{\rm rel}\sigma_{\rm Yb Rb}\right>$ and $K_3$ are used to fit the shape and the lifetime of
the experimental data. A good fit is obtained for $K_3 = 1.1 \, \times \,
10^{-26} {\rm cm}^6$s$^{-1}$ and 
$\left<v_{\rm rel}\sigma_{\rm YbRb}\right>$ one fourth of the unitarity limit, corresponding to a
zero-energy scattering length of $|a_{\mathrm{YbRb}}| \approx 500\,a_0$ within
effective range theory.   A range of $\left<v_{\rm rel}\sigma_{\rm YbRb}\right>$ and $K_{3}$ give
similar profiles with lifetimes consistent with experiment. A more
detailed experimental investigation of the Rb density-dependence of the lifetime
together with full 3-D calculations are required to deduce definite values for
the collisional parameters, which goes beyond the scope of the present
manuscript.

As a conclusion, we have observed spatial separation in a thermal mixture of $^{87}$Rb and $^{174}$Yb driven by  large interspecies interactions.  
A  mean-field model cannot explain the observations quantitatively, although it suggests large interspecies interaction.  A diffusive model including  three-body recombination can reproduce the experimental observations with physically reasonable parameters. To resolve the exact physical origin of the spatial separation, further experimental and theoretical investigations in a simplified trapping geometry are planned.
The  evidence of strong interactions in this system together with measurements of thermalization rates for other Yb isotopes has already been used to predict collisional properties and binding energies near threshold  for all isotopic combinations of Yb and Rb \cite{maxwell2010a} .  

The project is supported from DFG under 
SPP 1116. F.B. was supported by the Stiftung der Deutschen Wirtschaft.


%
\end{document}